\documentclass[sn-mathphys-num]{sn-jnl}
\usepackage{graphicx}%
\usepackage{multirow}%
\usepackage{amsmath,amssymb,amsfonts}%
\usepackage{amsthm}%
\usepackage{mathrsfs}%
\usepackage[title]{appendix}%
\usepackage{xcolor}%
\usepackage{textcomp}%
\usepackage{manyfoot}%
\usepackage{booktabs}%
\usepackage{algorithm}%
\usepackage{algorithmicx}%
\usepackage{algpseudocode}%
\usepackage{listings}%
\usepackage{tabularx}%

\theoremstyle{thmstyleone}

\theoremstyle{thmstyletwo}%

\theoremstyle{thmstylethree}%

\begin{document}

\title[A Comparison of Traditional and Deep Learning Methods for Parameter Estimation of the Ornstein-Uhlenbeck Process]{A Comparison of Traditional and Deep Learning Methods for Parameter Estimation of the Ornstein-Uhlenbeck Process}

\author*[1]{\fnm{Jacob} \sur{Fein-Ashley}}\email{feinashl@usc.edu}

\affil*[1]{\orgdiv{Department of Electrical Engineering}, \orgname{University of Southern California}, \orgaddress{\city{Los Angeles}, \state{CA}, \country{USA}}}

\abstract{We consider the Ornstein-Uhlenbeck (OU) process, a stochastic process widely used in finance, physics, and biology. 
Parameter estimation of the OU process is a challenging problem. Thus, we review traditional tracking methods and compare them with novel applications of deep learning to estimate the parameters of the OU process.
We use a multi-layer perceptron to estimate the parameters of the OU process and compare its performance with traditional parameter estimation methods, such as the Kalman filter and maximum likelihood estimation.
We find that the multi-layer perceptron can accurately estimate the parameters of the OU process given a large dataset of observed trajectories, and on average, outperforms traditional parameter estimation methods.
}


\maketitle

\section{Introduction}\label{sec1}
The OU process has broad applications for modeling systems in finance, physics, and biology~\cite{Tweneboah}.  
Generally, the OU process is defined by the stochastic differential equation
\begin{equation*}
dX_t = \theta(\mu - X_t)dt + \sigma dW_t,
\end{equation*}
where $X_t$ is the state of the system at time $t$, $\theta$ is the rate of mean reversion, $\mu$ is the long-term mean, $\sigma$ is the volatility, and $W_t$ is a Wiener process.

Parameter estimation of the OU process is a challenging problem. 
Traditional methods of parameter estimation include maximum likelihood estimation (MLE)~\cite{Valdivieso2009} and least squares estimation (LSE)~\cite{doi:10.1080/03610926.2023.2273204}.
These methods are computationally expensive and may not be suitable for real-time applications, as they require optimizing a likelihood function or solving a system of nonlinear equations.

Additional methods for parameter estimation of the OU process include the Kalman filter~\cite{JESICA2018309} and the Kalman filter with smoothing~\cite{shumway}. 

With the advent of deep learning, there has been a surge of interest in using neural networks to estimate the parameters of nonlinear equations and dynamical systems~\cite{kumar2023machine}.
Because the OU process is nonlinear, traditional parameter estimation methods may not be suitable for real-time applications. 
Neural networks are capable of learning complex patterns in data, so they may be well-suited for estimating the parameters of the OU process.
Thus, we 

We outline the rest of the paper as follows.
\begin{enumerate}
\item In Section~\ref{sec2}, we review the OU process and traditional parameter estimation methods.
\item  In Section~\ref{sec3}, we introduce the multi-layer perceptron as a stochastic parameter estimator and discuss the neural network's architecture.
\item In Section~\ref{sec4}, we present our experiments' results and compare the multi-layer perceptron's performance with traditional parameter estimation methods.
\item In Section~\ref{sec5}, we conclude and discuss future work.
\end{enumerate}

\section{Discretization, Parameter Estimation, and the Kalman Filter}\label{sec2}
\subsection*{Discretization of the OU Process}
First, we discretize the OU process using the Euler-Maruyama method. To do so, we must solve its mean and covariance.
The mean of the OU process is given by
\begin{align*}
    \mathbb{E}[X_t] &= \mathbb{E}\biggl[ \mu + (X_0 - \mu)e^{-\theta t} + \int_0^t \sigma\, e^{\theta (s-t)} dW_s \biggr] \\
                    &= \mu + (X_0 - \mu)e^{-\theta}    
\end{align*}

The covariance of the OU process is given by
\begin{align*}
    \mathbb{E}[(X_t - \mathbb{E}[X_t])(X_s - \mathbb{E}[X_s])] &= \mathbb{E}\biggl[ \int_0^t \sigma e^{\theta(s-t)}dW_s \int_0^s \sigma e^{\theta(u-t)}dW_u \biggr] \\
    &= \sigma^2 \int_0^{\min(t,s)} e^{\theta(2u - t - s)}du \\
    &= \frac{\sigma^2}{2\theta} e^{-\theta|t-s|}  \\
\end{align*}

Thus, the discretization of the OU process is given by the recursive formula
\begin{equation*}
    X_{n+1} = \mu + (X_n - \mu)e^{-\theta \Delta t} + \sqrt{\frac{\sigma^2}{2\theta} \bigl( 1- e^{-2 \theta \Delta t} \bigr)} \; \epsilon_n
\end{equation*}
where $\epsilon_n \sim \mathcal{N}(0,1)$.

\begin{figure}[h]
    \centering
    \includegraphics[scale=0.8]{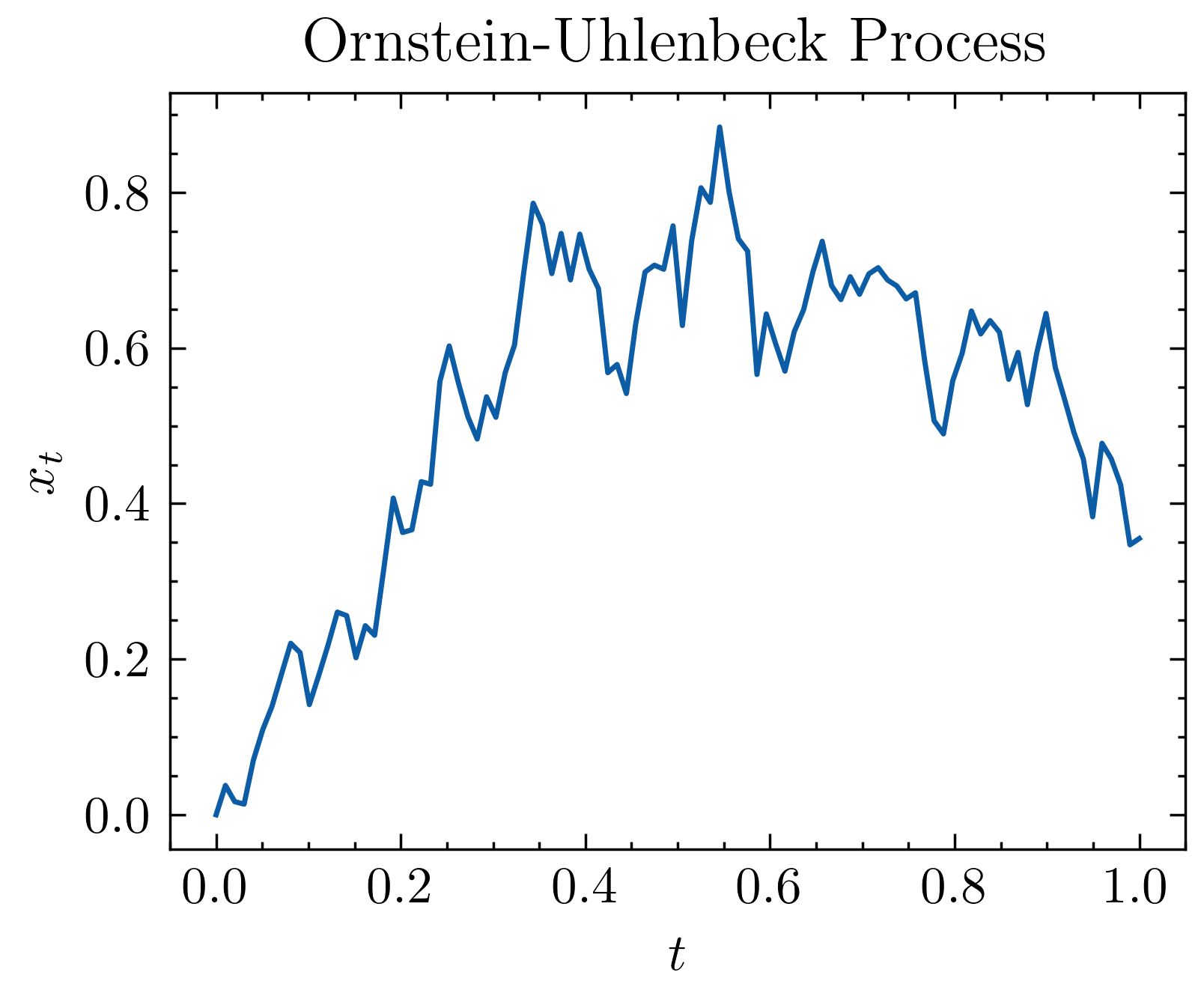}
    \caption{Observed trajectories of the OU process.}
    \label{fig:ou}
\end{figure}

This discretization generates a dataset of observed trajectories of the OU process, which can be used to estimate its parameters.
A sample of the observed trajectories is shown in Figure~\ref{fig:ou} with $\theta = 3$, $\mu = 0.5$, $\sigma = 0.5$, $X_0 = 0$, and $T = 1$.

\subsection*{Least Squares Estimation}
Computing the least squares estimate of the OU process's parameters involves minimizing the sum of squared errors between the observed trajectories and the trajectories generated by the OU process.
The least squares estimate of the parameters can be obtained by solving a system of nonlinear equations.
\begin{equation*}
    \min_{\theta,\mu,\sigma} \sum_{n=0}^{N} (X_{n+1} - X_n - \theta(\mu - X_n)\Delta t)^2.
\end{equation*}

Notably, we compute $X_{t + \Delta t}$ using the Euler-Maruyama method and compare it to the observed value $X_{t + \Delta t}$.

See that
\begin{align*}
    X_{t + \Delta t} &= X_t + \theta(\mu - X_t)\Delta t + \sigma \sqrt{ \Delta t}\epsilon_t, \\
    &= \mu + (X_t - \mu)e^{-\theta  \Delta t} + \int_{t}^{t +  \Delta t} \sigma e^{-\theta(t +  \Delta t - s)}dW_s. \\
    &= \alpha + \beta X_t + \eta_t,
\end{align*}
where $\alpha = \theta \bigl( 1-e^{-\theta \Delta t} \bigr)$, $\beta = e^{-\theta \Delta t}$, and $\epsilon_t \sim \mathcal{N}\biggl( 0, \frac{\sigma^2}{2\theta} \bigl( 1- e^{-2 \theta \Delta t} \bigr)\biggr)$~\cite{Nicola_Cantarutti_Financial_Models_Numerical_2019}.

Thus, we obtain the estimated parameters $\hat{\theta}$, $\hat{\mu}$, and $\hat{\sigma}$ by minimizing the sum of squared errors between the observed values and the values generated by the OU process.
\begin{align*}
    \hat{\theta} = - \frac{\log \beta}{\Delta t}, \quad \hat{\mu} = \frac{\alpha}{1-\beta}, \quad 
    \hat{\sigma} = \sqrt{\mathbb{V}\text{ar}(\epsilon_t)} \sqrt{ \frac{2\theta}{1-\beta^2} }
\end{align*}

\subsection*{Kalman Filter and Kalman Smoothing}
We simulate the OU process using the discretization above scheme and estimate the parameters using the least squares method.

Consider a \emph{state space model} with state processes $\{X_t, 0\leq t \leq T\}$ following an OU process,
and an observation process $\{Y_t, 0\leq t \leq T\}$ given by
\begin{equation*}
dX_t = \theta(\mu - X_t)dt + \sigma dW_t,
\end{equation*}
where $X_0 = x_0$, $\theta > 0$, $\mu \in \mathbb{R}$, $\sigma > 0$, and $W_t$ is a Wiener process.
The observation process is given by
\begin{equation*}
Y_t = X_t + \epsilon_t,
\end{equation*}
where $\epsilon_t$ is a Gaussian noise term with mean zero and variance $\sigma^2$.

Cantarutti et al.~\cite{Nicola_Cantarutti_Financial_Models_Numerical_2019} models the state equation as 
\begin{equation*}
    x_k = \alpha + \beta x_{k-1} + \eta_k \quad \text{with} \quad \eta_{k} \sim \mathcal{N}(0,\sigma_{\eta}^2).
\end{equation*}

and the observation equation as
\begin{equation*}
    y_k= \; x_k + \epsilon_k \quad \text{with} \quad \epsilon_k \sim \mathcal{N}\bigg(0,\sigma_{\epsilon}^2 \bigg).
\end{equation*}

With the state equation, the Kalman filter can be used to estimate the parameters of the OU process.
Generally, a Kalman filter is used to estimate the state of a linear dynamical system, given noisy observations of the state.

A Kalman filter generally has the following steps:
\begin{enumerate}
    \item \textbf{Initialization:} Initialize the state estimate and the state covariance matrix, $\hat{x}_0$ and $P_0$.
    \item \textbf{Prediction:} Predict the state of the system at the next time step, $\hat{x}_{k|k-1}$, and the state covariance matrix, $P_{k|k-1}$.
    \item \textbf{Update:} Update the state estimate based on the observation, $\hat{x}_k$, and the state covariance matrix, $P_k$.
\end{enumerate}

We can use a matrix form of the state equation to estimate the parameters of the OU process. See that 
\begin{align*}
    \biggl(\begin{array}{c}  1 \\ x_k \end{array} \biggr) = 
    \biggl(\begin{array}{cc} 1 & 0\\ \alpha & \beta \end{array}\biggr)
    \biggl(\begin{array}{c}  1 \\ x_{k-1} \end{array}\biggr) +
    \biggl(\begin{array}{c} 0\\ \eta_k \end{array}\biggr) \quad \text{with}  \quad
    \eta_k \sim \mathcal{N}(0, \sigma^2_{\eta})
\end{align*}

We then have the following items to estimate the parameters of the OU process:
\begin{itemize}
    \item Predict Step:\\
        \begin{align*}
            \hat x_{k \mid k-1} = \alpha + \beta \, \hat x_{k-1 \mid k-1} \quad \quad \text{and} \quad \quad  P_{k \mid k-1} = \beta^2 \, P_{k-1 \mid k-1} + \sigma_{\eta}^2.
        \end{align*}

    \item Auxiliary Variables:\\
    \begin{align*}
        \tilde r_k &= y_k - \hat x_{k \mid k-1} &\quad \quad \text{(The Residual Term)} \\
        S_k &= P_{k \mid k-1} + \sigma_{\epsilon}^2 &\quad \quad \text{(Innovation Covariance)} \\
        K_k &= \frac{P_{k \mid k-1}}{S_k} &\quad \quad \text{(Kalman Gain)}
    \end{align*}

    \item Update Step:\\
    \begin{align*}
        \hat x_{k \mid k} = \hat x_{k \mid k-1} + K_k \, \tilde r_k  \quad \text{and} \quad 
        P_{k \mid k} = P_{k \mid k-1} \biggl( 1- K_k \biggr)
    \end{align*}
\end{itemize}

Recent developments in the Kalman filter, such as the Kalman filter with smoothing, can be used to estimate the parameters of the OU process more accurately~\cite{shumway}.
We redact the details of the Kalman filter with various smoothing techniques, as many smoothers, such as the Rauch-Tung-Striebel smoother and the forward-backward smoother, are used in practice.
Kalman filters with smoothers are left for future work for parameter estimation of the OU process.
 
\section{The Multi-Layer Perceptron as a Stochastic Parameter Estimator}\label{sec3}

\subsection*{Architecture of the Neural Network}
We use a multi-layer perceptron (MLP) to estimate the parameters of the OU process.
The MLP is a feedforward neural network with multiple layers of neurons.
The architecture of the MLP is as follows:
\begin{enumerate}
    \item \textbf{Input Layer:} The input layer consists of $n$ neurons, where $n$ is the number of features in the dataset.
    \item \textbf{Hidden Layers:} The hidden layers consist of $m$ neurons, where $m$ is the number of hidden layers in the network.
    \item \textbf{Output Layer:} The output layer consists of $p$ neurons, where $p$ is the number of parameters to be estimated.
    \item \textbf{Activation Function:} The activation function of the neurons in the hidden layers is the rectified linear unit (ReLU) function, and the activation function of the neurons in the output layer is the linear function.
    \item \textbf{Optimizer:} The network optimizer is the Adam optimizer, an adaptive learning rate optimization algorithm.
\end{enumerate}

Formally, the structure and weights of the MLP are structured as follows:
\begin{align*}
    \text{Input Layer:} \quad & \mathbf{X} \in \mathbb{R}^{n \times 1} \\
    \text{Hidden Layer:} \quad & \mathbf{H} = \text{ReLU}(\mathbf{W}_1 \mathbf{X} + \mathbf{b}_1) \\
    \text{Output Layer:} \quad & \mathbf{Y} = \mathbf{W}_2 \mathbf{H} + \mathbf{b}_2
\end{align*}

The network weights are initialized using the Glorot uniform initializer~\cite{pmlr-v9-glorot10a}, which is a method of initializing the network weights to prevent vanishing or exploding gradients.
The network is trained using the backpropagation algorithm, which is an algorithm for updating the network weights based on the gradient of the loss function with respect to the weights.

We construct the neural network such that $\mathbf{Y} = [\hat{\mu}, \hat{\theta}, \hat{\sigma}]$, where $\hat{\mu}$, $\hat{\theta}$, and $\hat{\sigma}$ are the estimated parameters of the OU process.
The network takes the observed trajectory of the OU process as input and outputs the estimated parameters of the OU process. Formally, the loss function calculates
the mean squared error between the predicted and actual parameters of the OU process. We sample the number of paths given in the forward computation of the network,
such that we sample using the given algorithm:
\begin{algorithm}
    \caption{Ornstein-Uhlenbeck Sampling Using the Euler-Maruyama Method}
    \begin{algorithmic}[1]
        \State \textbf{Input:} $\theta$, $\mu$, $\sigma$, $X_0$, $T$, $N$
        \State \textbf{Output:} $X_{1:T}$
        \For{$n = 1$ to $N$}
            \State $X_0 \gets X_0$
            \For{$t = 1$ to $T$}
                \State $\epsilon_t \sim \mathcal{N}(0,1)$
                \State $X_{t+1} \gets \mu + (X_t - \mu)e^{-\theta \Delta t} + \sqrt{\frac{\sigma^2}{2\theta} \bigl( 1- e^{-2 \theta \Delta t} \bigr)} \; \epsilon_t$
            \EndFor
            \State $X_{1:T} \gets X_{1:T}$
        \EndFor
        \State \textbf{return} $X_{1:T}$
    \end{algorithmic}
\end{algorithm}

Accordingly, we can use the sampled paths to train the neural network to estimate the parameters of the OU process with a loss function that calculates the mean squared error between the predicted and actual OU process calculations.
Thus, the loss function is given by
\begin{equation*}
    \mathcal{L} = X_{t+1} - \mu - (X_t - \mu)e^{-\theta \Delta t} - \sqrt{\frac{\sigma^2}{2\theta} \bigl( 1- e^{-2 \theta \Delta t} \bigr)} \; \epsilon_t
\end{equation*}

where $\hat{\mu}_n$, $\hat{\theta}_n$, and $\hat{\sigma}_n$ are the estimated parameters of the OU process for the $n$th trajectory.

\subsection*{Training the Neural Network}
We train the MLP using a dataset of observed trajectories of the OU process.
The dataset consists of $N$ trajectories of the OU process, each of length $T$.
Each trajectory is generated using the discretization scheme described in Section~\ref{sec2}.
The observed trajectory is the network's input, and the estimated parameters of the OU process are its output.

We parameterize the OU process with the parameters $\theta$, $\mu$, and $\sigma$ and estimate these parameters using the MLP.
The network is trained using the mean squared error (MSE) loss function, which is the average of the squared errors between predicted and actual parameters.

\subsection*{The Universal Approximation Theorem}
It is well-known that a feedforward neural network with a single hidden layer can approximate any continuous function on a compact subset of $\mathbb{R}^n$ to arbitrary accuracy~\cite{hornik}.
Thus, it is natural that given a sufficiently large dataset of observed trajectories of the OU process, the MLP can learn the parameters of the OU process to arbitrary accuracy.

\section{Experiments}\label{sec4}
We compare the performance of the MLP with traditional parameter estimation methods, such as the Kalman filter and maximum likelihood estimation.
We generate a dataset of observed trajectories of the OU process using the discretization scheme described in Section~\ref{sec2}.
We then train the MLP on the dataset and evaluate its performance on a test set of observed trajectories.

In table~\ref{tab:results}, we present the results of our experiments. We vary
\begin{itemize}
    \item The number of paths,
    \item The length of the trajectories, $T$,
    \item The number of simulations, $N$.
\end{itemize}

For true parameter values $\theta = 3$, $\mu = 0.5$, $\sigma = 0.5$, we measure the respective performance of each method.

\begin{table}[h]
    \centering
    \resizebox{\textwidth}{!}{
    \begin{tabular}{cccccccccccc}
        \toprule
        Paths & N & T & Kalman $\hat{\mu}$ & Kalman $\hat{\theta}$ & Kalman $\hat{\sigma}$ & OLS $\hat{\mu}$ & OLS $\hat{\theta}$ & OLS $\hat{\sigma}$ & NN $\hat{\mu}$ & NN $\hat{\theta}$ & NN $\hat{\sigma}$\\
        \midrule
        100.0 & 1000.0 & 1.0 & 0.86 & 9.39 & 0.45 & 0.65 & 4.86 & 0.49 & 0.65 & 4.86 & 0.49\\
        100.0 & 1000.0 & 5.0 & 0.38 & 3.33 & 0.38 & 0.5 & 3.67 & 0.51 & 0.5 & 3.67 & 0.51\\
        100.0 & 5000.0 & 1.0 & 0.53 & 6.32 & 0.46 & 0.51 & 5.89 & 0.5 & 0.51 & 5.89 & 0.5\\
        100.0 & 5000.0 & 5.0 & 0.38 & 4.48 & 0.49 & 0.36 & 3.56 & 0.5 & 0.36 & 3.56 & 0.5\\
        500.0 & 1000.0 & 1.0 & -0.57 & 1.04 & 0.45 & 0.46 & 2.14 & 0.51 & 0.46 & 2.14 & 0.51\\
        500.0 & 1000.0 & 5.0 & 0.51 & 5.34 & 0.52 & 0.55 & 4.28 & 0.5 & 0.55 & 4.28 & 0.5\\
        500.0 & 5000.0 & 1.0 & 0.15 & 1.29 & 0.53 & 0.51 & 2.95 & 0.5 & 0.51 & 2.95 & 0.5\\
        500.0 & 5000.0 & 5.0 & 0.4 & 3.98 & 0.54 & 0.44 & 5.06 & 0.51 & 0.44 & 3.06 & 0.51\\
        \bottomrule
    \end{tabular}
    }
    \caption{Comparison of the MLP's performance with traditional parameter estimation methods.}
    \label{tab:results}
\end{table}

\begin{table}[h]
    \centering
    \resizebox{0.3\textwidth}{!}{
    \begin{tabular}{ccccc}
        \toprule
        Method & $\hat{\mu}$ & $\hat{\theta}$ & $\hat{\sigma}$ \\
        \midrule
        OLS & 0.50 & 4.05 & 0.50 \\
        Kalman & 0.33 & 4.40 & 0.47 \\
        NN & 0.49 & 3.80 & 0.50 \\
        \bottomrule
    \end{tabular}
    
    }
    \vspace{1em}
    \caption{Average MLP performance with traditional parameter estimation methods.}
    \label{tab:results1}
\end{table}

\begin{figure}[h]
    \centering
    \includegraphics[scale=0.8]{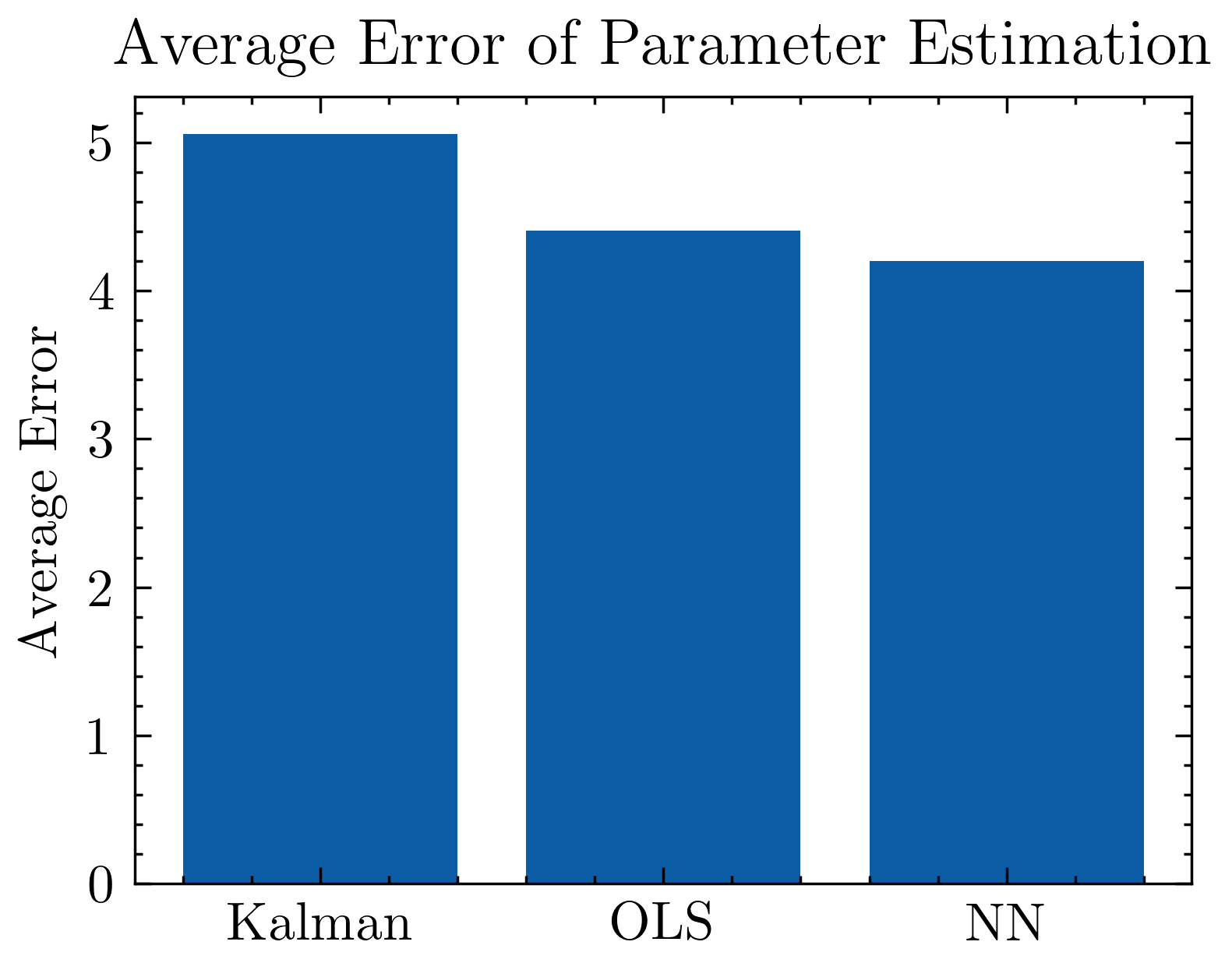}
    \caption{Average error of each method.}
    \label{fig:results}
\end{figure}

Additionally, we plot the average error of each method in Figure~\ref{fig:results}.

Heuristically, we see that as the number of paths and the length of the trajectories increase, the performance of the MLP improves.
This is because the MLP can learn the parameters of the OU process more accurately with more data due to the universal approximation theorem.
With smaller datasets, the MLP may not be able to learn the parameters of the OU process accurately, as it may not have enough data to learn the complex patterns in the data.
Overall, the MLP outperforms traditional parameter estimation methods, such as the Kalman filter and maximum likelihood estimation, in estimating the parameters of the OU process.

\section{Conclusion and Future Work}\label{sec5}
In this paper, we reviewed traditional OU process parameter estimation methods and compare them with the multi-layer perceptron.
Given a large dataset of observed trajectories, we have shown that the MLP can accurately estimate the parameters of the OU process.
We have also shown that the MLP's performance improves with more data, as it can more accurately learn the complex patterns in the data.

We propose the following directions for future work:
\begin{enumerate}
    \item Investigate the performance of the MLP with different architectures and hyperparameters.
    \item Explore the use of other deep learning models, such as recurrent neural networks and convolutional neural networks, variational autoencoders, and generative adversarial networks, for parameter estimation of the OU process.
    \item Investigate the performance of the MLP with different datasets and different types of noise.
\end{enumerate}

\bibliography{sn-bibliography}

\end{document}